# Longitudinal Sentiment Topic Modelling of Reddit Posts

Fabian Nwaoha    Ziyad Gaffar     Ho Joon Chun    Marina Sokolova

University of Ottawa

{fnwao011, cazgaff021, hchun043, sokolova}@uottawa.ca

**Abstract.** In this study, we analyze texts of Reddit posts written by students of four major Canadian universities. We gauge the emotional tone and uncover prevailing themes and discussions through longitudinal topic modeling of posts' textual data. Our study focuses on four years, 2020-2023, covering COVID-19 pandemic and after pandemic years. Our results highlight a gradual uptick in discussions related to mental health.

## 1 Introduction

Our current study focuses on the topic modelling of sentiments expressed by university students in their online Reddit posts. Surveys of this age demographics had shown an elevated level of anxiety during the COVID-19 pandemic, e.g., National College Health Assessment (Abdi, 2023), the Household Pulse Survey (Panchal et al., 2021). The above studies did not include topic modelling of texts posted on social networks favored by the student population. We also extend the study by including years after the COVID pandemic. We work with posts from students of four universities: Waterloo, University of Toronto (UofT), McGill, University of British Columbia (UBC).

Reddit is one of the most popular social media platforms, having over 52 million daily active users (Backlinko, 2023). The platform is built around topic-specific subreddits that enable users to engage in discussions of niche subjects. Notably, subreddits can be university-specific, offering a platform for students from individual institutions to engage in focused discussions. As Reddit can be used anonymously, it encourages users to have more open and candid interactions.

Sawicki et al (2021) show a high usability of the Reddit data for topic analysis, information processing, linguistic analysis, etc. Moreover, the ability to categorize data through different subreddits enhances the potential for nuanced analysis of various demographics (Yang et al, 2022) The authors analyze positive and sentiments in COVID-19-related messages from subreddits r/Canada and r/Unitedkingdom. Their technique uses sentiment lexicons VADER and TextBlob for sentiment gauging and the data attrition for convergence of manual sentiment annotations and improvement of sentiment classification.

Other related studies include (Xie et al,2020), with the focus on public response to COVID-19 on Weibo, a Chinese microblog platform, (Qi & Shabrina, 2023) that compares the effectiveness of lexicon-based and machine learning methods for sentiment analysis of Twitter data and presents a comparative study of different lexicon combinations, while also exploring the importance of feature generation and selection processes for machine learning sentiment classification, and (Jelodar et al 2020) that employs NLP techniques and LSTM recurrent neural networks to analyze public sentiment and uncover topics of COVID-19 discussions on social media.

Longitudinal sentiment analysis of Twitter data reported on the use of emoticons and emojis without delving into linguistic aspects of expressed sentiments (Yin et al, 2021). Longitudinal studies of Reddit posts mostly analyzed mental health and substance use (Alambo et al, 2021; Yang et al, 2023).

Our study differs from the above in that i) we employ advanced analytics tools for in-depth text studies; ii) we use four year-long time intervals: 2020, 2021, 2022, and 2023.

The current study continues and is built upon our previous results reported in Nwahoa et al (2024).

## 2 Study Background

We model topics in texts from Reddit posts from four year-long intervals: 2020, 2021, 2022, and 2023. Those years cover the peak of COVID-19 pandemic and after-pandemic years. In respect to the university students in Canada, this means a shift to online learning in March 2020 – Aug 2021, then hybrid, online and in class, leaning in Sept 2021 – Aug 2022, and the full return to classes in Sept 2022. As the first instance of COVID-19 emerged in January 2020, our investigation encompasses the entire period spanning from 2020 to 2022. Given the commencement date of our research, our emphasis is on data gathered from January to November 2023.

We collected data from subreddits of four prominent Canadian universities: Waterloo[1], McGill[2], UBC[3], and UofT[4]. UofT, McGill, and UBC are the top three research universities in Canada[5]. Waterloo holds #1 in reputational survey and is constantly ranked among the top three comprehensive universities (universities that do not have medical schools)[6].

We have utilized several advanced tools to effectively analyze and interpret the sentiments of texts extracted from Reddit posts. Through application of LDA (Latent Dirichlet Allocation), a generative probabilistic model of a corpus, we can find common themes and relationships between words. The basic idea is that documents are represented as random mixtures over latent topics, where each topic is characterized by a distribution over words (Blei et al, 2002). We used implementations by Gensim[7], a Python library, that allows us to optimize the LDA models by calculating coherence scores and finding the optimal number of topics.

We employed VADER[8] (Valence Aware Dictionary and Sentiment Reasoner) to gauge sentiments expressed in the posts. It allowed us to calculate the presence of negative and positive words. With VADER, we delved into the sentiments of different topics found through LDA and identified the key words associated with their sentiments. TextBlob[9] computes polarity scores, which can be crucial in understanding the sentimental context of the post. Utilizing TextBlob, we aimed to identify positive and negative sentiments, extracting keywords that carried significant sentimental weight.

## 3 Data Set Construction

**Data preprocessing.** We employed keywords related to mental health, such as *stress, worried, nervous, uneasy, distress, fear, concern, unmotivated, anxious, anxiety, unease,* and *mental health*. Figure 1 shows which keywords yielded the most substantial number of entries and held greater prevalence in the text. Next, we decided to clean the texts by removing stopwords ('there', 'the', 'is' are examples of stopwords) This step eliminates any commonly occurred words that carry little to no significance in determining sentiment, which leads to the minimization of noise in the dataset. This assists in improving the accuracy and effectiveness of sentiment analysis algorithms in understanding and categorizing sentiments accurately.

We also dropped blank comments that occurred in the data. This led to a decrease in the overall entries from 4,507 to 4,205. Initially, we chose to employ the original dataset with 4,507 entries at the onset of applying sentiment evaluation to the features, aiming to discern performance variations between the two sentiment analysis tools. Subsequently, for the dataset containing 4,205 entries, we employed it specifically during sentiment classification. This decision was driven by the need to eliminate blank comments to reduce potential noise, enhancing machine learning performance and outcomes. Additionally, we utilized this dataset for determining the optimal topic modeling, as blank comments held no relevance in that context.

---

[1] https://www.reddit.com/r/uwaterloo/
[2] https://www.reddit.com/r/mcgill/
[3] https://www.reddit.com/r/UBC/
[4] https://www.reddit.com/r/UofT
[5] Canada's Best Medical Doctoral Universities: Rankings 2024 | Maclean's Education (macleans.ca)
[6] Canada's Best Comprehensive Universities: Rankings 2024 | Maclean's Education (macleans.ca)
[7] https://pypi.org/project/gensim/
[8] https://github.com/cjhutto/vaderSentiment
[9] https://textblob.readthedocs.io/en/dev/

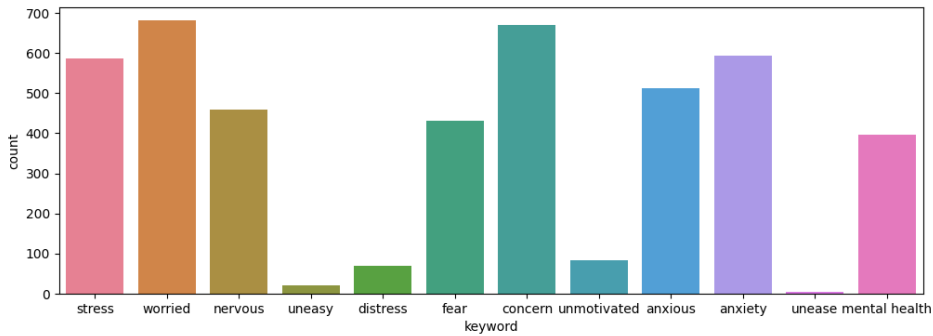

Figure 1: Occurrence of keywords in the data set.

As a result, we obtained four relatively balanced data sets. Figures 2 and 3 report post distribution among the universities for each year.

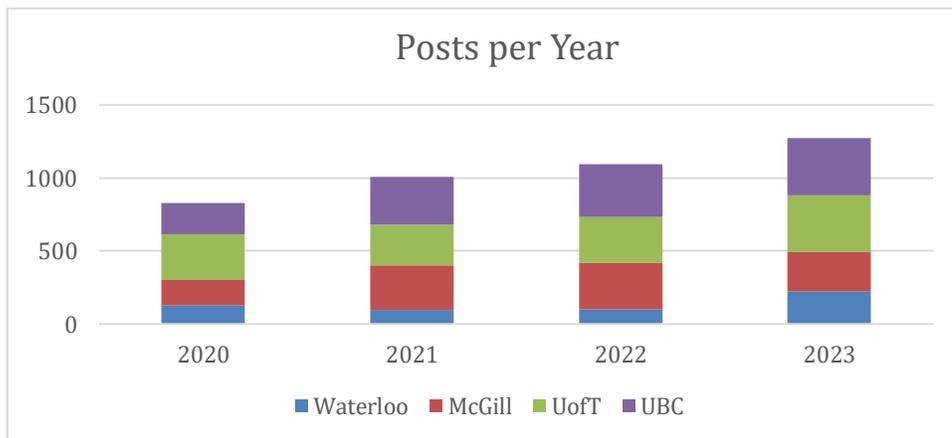

Figure 2: Number of extracted Reddit posts per year.

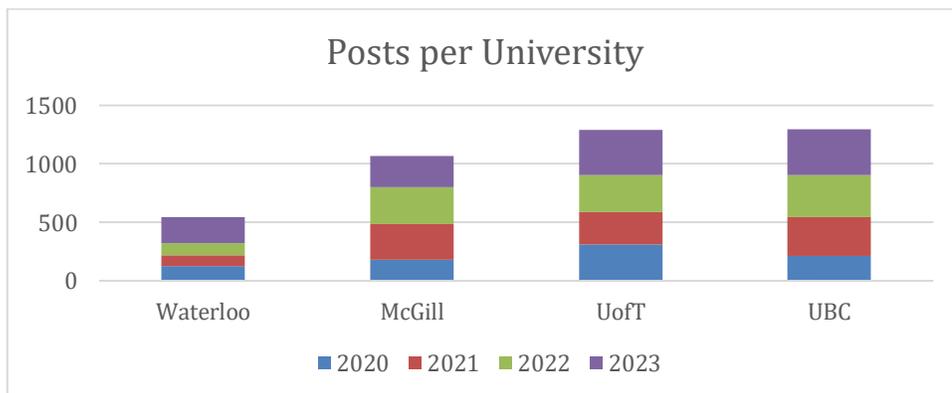

Figure 3: Number of extracted Reddit posts per university

Sentiment evaluation of the texts goes as follows:

1. We compute both VADER and TextBlob polarity scores for each cleaned text body, which is then placed under a column called 'Vader Sentiment' or 'Textblob Sentiment'.
2. We decide on a threshold value that would be in place to filter out whether the body text or the word is a positive or negative sentiment.
3. After obtaining the best threshold value (more explained under the 'Threshold Value'), utilize it to determine if a text is positive, negative, or neutral.
4. While still using the same threshold value to determine under which sentiment it belongs, we extract the wordlist of each of the sentiment polarities (positive, negative, neutral) by running each word in the cleaned text body through both sentiment tools (VADER and Textblob).
5. Count the total number of words within each of the sentiment wordlist, then placing it under a new column that is specifically for each sentiment polarity (positive, negative, neutral).

## 4 Optimal Topic Modelling

Determining the optimal number of topics for an LDA model tailored to a specific dataset is a crucial aspect of topic modeling. Due to the absence of a one-size-fits-all solution, it is important to employ a systematic approach in assessing the coherence score for each model across various topic numbers. To achieve this, we leverage the four-stage topic coherence framework, encompassing segmentation, probability calculation, confirmation measure, and aggregation, using the Gensim library (Röder et al, 2015). We construct multiple LDA($i$) models, where $i$ represents the number of topics under consideration. Throughout this iterative process, the sole parameter that undergoes modification is the number of topics, keeping all other factors constant. Figure 4 illustrates computation of coherence scores, a pivotal step in the evaluation process. By systematically adjusting the number of topics, we aim to find the optimal configuration for the LDA model in the context of our specific dataset. The coherence score reached its peak when the number of topics was set to 7. As a result, we made the decision to use 7 topics as the parameter throughout the project. This choice ensures that our topic modeling strategy is fine-tuned to the inherent characteristics of the data.

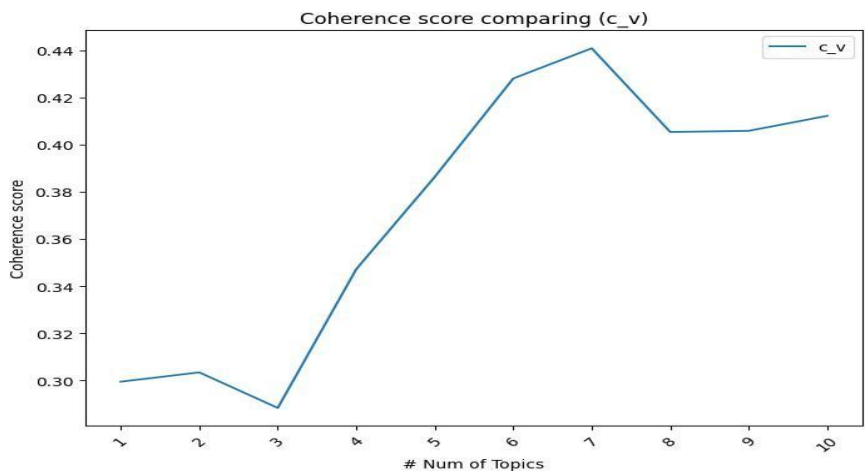

*Figure 4: Coherence Score for the number of topics*

### 4.3 Topic Modelling results

First, we constructed an LDA model and applied it to the complete dataset (Table 1). LDA (Latent Dirichlet Allocation) yields relevant topic words, but the task of formulating an overarching topic or theme requires human review and interpretation. To achieve this, we employ a reader-centric approach (Strapparava and Mihalcea, 2008), where topic labels are assigned based on reader perception. Three authors (FN, ZG, HJH) agree on the topic labels.

*Table 1: Topic modelling results for the entire data set.*

| | | |
|---|---|---|
| Topic 0 | '0.031*"exam" + 0.028*"class" + 0.028*"course" + 0.023*"final" + ' '0.020*"exams" + 0.014*"questions" + 0.012*"prof" + 0.011*"midterm" + ' '0.010*"study" + 0.010*"assignments"' | University |
| Topic 1 | 0.034*"ams" + 0.030*"mar" + 0.025*"coverage" + 0.019*"appointment" + ' '0.019*"club" + 0.016*"doctor" + 0.014*"diagnosis" + 0.011*"morning" + ' '0.010*"folks" + 0.009*"doctors"' | Health care |
| Topic 2 | '0.019*"study" + 0.019*"program" + 0.012*"research" + 0.011*"offer" + ' '0.011*"arts" + 0.010*"access" + 0.010*"lab" + 0.009*"receive" + ' '0.009*"grad" + 0.008*"email"' | University |
| Topic 3 | '0.020*"participate" + 0.020*"httpspreview" + 0.016*"experiencing" + '0.016*"pngwidth" + 0.012*"selvesubc" + 0.011*"admission" + 0.010*"jpgwidth" + 0.008*"accessing" + 0.007*"withdraw" + 0.007*"qualtrics"' | Undetermined |
| Topic 4 | '0.020*"feel" + 0.017*"people" + 0.013*"even" + 0.012*"anxiety" + 0.009*"go" ' '+ 0.008*"cant" + 0.008*"friends" + 0.007*"time" + 0.007*"things" + ' '0.007*"life"' | Anxiety |
| Topic 5 | '0.021*"year" + 0.017*"would" + 0.013*"first" + 0.012*"time" + 0.011*"also" ' '+ 0.011*"anyone" + 0.010*"school" + 0.009*"take" + 0.008*"well" + ' '0.008*"help"' | First year students |
| Topic 6 | '0.044*"health" + 0.040*"mental" + 0.030*"students" + 0.014*"please" + ' '0.011*"support" + 0.009*"us" + 0.009*"covid" + 0.008*"free" + ' '0.007*"services" + 0.007*"issues"' | Mental health |

Next, we assessed the emotional tone associated with each topic. To achieve this, we assigned each post to its respective topic and processed it through VADER. We computed the average sentiment for each topic (Figure 5). Those results prompted us to reassess our assumptions. For example, despite the presence of negative terms within a topic, it did not uniformly render the entire topic negative. This insight highlighted the complexity of sentiment analysis, particularly when using VADER, where the entire post was processed, which introduced heightened variability in sentiment. Having the singular negative topic (Topic 0) was surprising to us (Fig 5). In our initial dataset analysis, we learned that positive posts outnumbered negative ones according to VADER, a potential contributing factor to this observation.

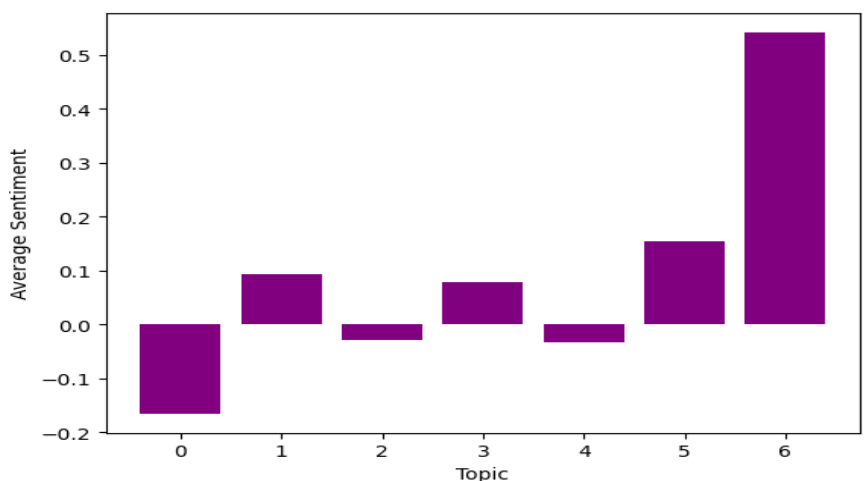

*Figure 5: Average sentiment per topic on the entire dataset.*

As we hypothesized that the data from these years would mirror real-life events, the subsequent phase of our studies involved a more detailed examination of the data, breaking it down by years. First, our focus will be on the years 2020 and 2021. Given that these were the years marked by the onset of the COVID-19 pandemic and the mandatory implementation of distance learning, our text analysis aimed to uncover insights related to either COVID-19 or virtual learning and their relationship with anxiety. In 2020 (Table 2), we noticed a considerable number of negative sentiments, but interestingly, only two topics aligned with our initial expectations. Topic 2 delves into the realm of teaching and learning, with keywords like learning, different, and profs hinting at a shift in learning methods, possibly reflecting the transition to online education. On the other hand, Topic 6 centers around COVID-19, featuring keywords such as coronavirus, support, and information. This topic captures discussions related to the pandemic, with individuals seeking support and information during the relatively unfamiliar circumstances of the year 2020.

To gain a more in-depth understanding of the text in the year 2020, we analyzed the most frequently occurring bi-grams and trigrams ; "bi-gram" refers to a sequence of two adjacent words, and "tri-gram" consists of three consecutive words. Despite the absence of COVID-related terms within these combinations, it is noteworthy that the bi-gram 'mental health' emerged a remarkable 247 times during this year, surpassing the frequency in any other year within our study. This high occurrence indicates a heightened focus on discussions related to mental health in 2020, likely attributed to the substantial changes and challenges brought about by the ongoing pandemic. Table 3 reports the results.

*Table 2: Topic modelling results for the year 2020.*

| Topic 0 | '0.062*"mental" + 0.061*"health" + 0.019*"issues" + 0.017*"therapy" + ' <br> '0.017*"access" + 0.017*"please" + 0.016*"talk" + 0.014*"resources" + ' <br> '0.014*"community" + 0.011*"others"' | Mental health |
|---|---|---|
| Topic 1 | '0.021*"services" + 0.016*"service" + 0.012*"students" + 0.010*"pop" + ' <br> '0.009*"grad" + 0.008*"huge" + 0.006*"personal" + 0.006*"helping" + "0.006*"training" + 0.006*"engineering"' | University |
| Topic 2 | '0.021*"students" + 0.013*"prof" + 0.011*"learning" + 0.011*"profs" + ' <br> '0.011*"people" + 0.009*"nbsp" + 0.008*"class" + 0.007*"mark" + 0.006*"show" ' <br> '+ 0.006*"different"' | Teaching |
| Topic 3 | 0.013*"seats" + 0.012*"space" + 0.011*"music" + 0.009*"group" + ' <br> '0.009*"abuse" + 0.009*"resource" + 0.008*"youd" + 0.007*"project" + ' <br> '0.007*"original" + 0.007*"members"' | Group work |
| Topic 4 | '0.012*"year" + 0.010*"also" + 0.008*"work" + 0.008*"school" + 0.008*"time" ' <br> '+ 0.008*"people" + 0.007*"would" + 0.006*"feel" + 0.006*"guys" + ' '0.006*"first"' | School life |
| Topic 5 | '0.012*"feel" + 0.011*"even" + 0.009*"one" + 0.009*"help" + 0.008*"time" + ' <br> '0.008*"would" + 0.008*"go" + 0.007*"anxiety" + 0.007*"going" + ' '0.007*"want"' | Anxiety |
| Topic 6 | "0.020*"mar" + 0.010*"students" + 0.009*"support" + 0.009*"free" + ' <br> '0.006*"name" + 0.006*"may" + 0.006*"bc" + 0.005*"coronavirus" + ' <br> '0.005*"information" + 0.005*"new"') | Covid |

*Table 3: 2020 common bi- and tri-grams.*

| Common Bi-grams | (('mental', 'health'), 247), (('pop', 'pop'), 181), (('first', 'year'), 85), (('even', 'though'), 55), (('anyone', 'else'), 48), (('university', 'life'), 39), (('anything', 'mind'), 38), (('similar', 'situations'), 37), (('please', 'keep'), 36), (('feel', 'free'), 35) |
|---|---|
| Common-Trigrams | [(('pop', 'pop', 'pop'), 180), (('welcome', 'mental', 'health'), 35), (('mental', 'health', 'monday'), 35), (('health', 'monday', 'use'), 35), (('monday', 'use', 'topic'), 35), (('use', 'topic', 'talk'), 35), (('topic', 'talk', 'anything'), 35), (('talk', 'anything', 'mind'), 35), (('anything', 'mind', 'regarding'), 35), (('mind', 'regarding', 'university'), 35)] |

In the year 2021, we observed a better alignment between the topics and our initial hypotheses (Table 4). Topic 0 incorporates keywords such as covid, campus, students, miss, and opinions. This topic likely revolves around the impact of COVID-19 leading to campus closures, and students expressing their longing for the campus environment or seeking information about the COVID-19 situation on campus. In Topic 3, the discussion primarily revolves around university-related matters, but the presence of the keyword 'respondus' indicates a focus on Respondus, an anti-cheating software used by the University of Ottawa for proctoring exams during virtual learning. Lastly, Topic 6 stands out as having the highest frequency of negative sentiments, evident from keywords like covid, s***, and life. This topic reflects discussions on how COVID-19 has adversely affected people's lives.

Bigrams and trigrams in the 2021 data provide us with deeper insights into discussions prevalent on the subreddits (Table 5). Notably, the bigram 'mental health' continues to stand out as a dominant theme, ranking as the top bigram occurrence. Additionally, the presence of 'social anxiety' indicates concerns about interacting with others due to the impact of COVID-19.

Trigrams reveal phrases such as 'mental health issues,' 'serious case covid,' 'worried catching covid,' and 'anyone else feeling.' These trigrams, marked by a notably negative tone, consistently appear as top occurrences. This underscores that in 2021, COVID-19 emerged as a significant concern for students, contributing to severe mental health issues and heightened anxiety.

*Table 4: The 2021 data topics*

| Topic 0 | 0.022*"campus" + 0.017*"covid" + 0.015*"students" + 0.012*"likely" + '<br>'0.012*"vancouver" + 0.011*"university" + 0.009*"curious" + 0.009*"opinions" '<br>'+ 0.007*"miss" + 0.007*"live"'), | Covid |
|---|---|---|
| Topic 1 | 0.023*"event" + 0.015*"google" + 0.011*"facebook" + 0.010*"pm" + '<br>'0.010*"hosting" + 0.010*"councillors" + 0.009*"dr" + 0.008*"older" + '<br>'0.007*"comdocumentd" + 0.007*"voice"' | Social event |
| Topic 2 | 0.013*"feel" + 0.013*"time" + 0.010*"even" + 0.010*"going" + 0.008*"year" + '<br>'0.008*"want" + 0.008*"would" + 0.008*"people" + 0.007*"anxiety" + '<br>'0.007*"classes"' | Anxiety |
| Topic 3 | '0.027*"please" + 0.016*"interested" + 0.014*"discussion" + '<br>'0.013*"programubc" + 0.013*"respondus" + 0.010*"analysis" + 0.010*"acam" + '<br>'0.009*"admission" + 0.009*"ca" + 0.009*"psychology"' | University |
| Topic 4 | "0.012*"year" + 0.010*"anyone" + 0.010*"course" + 0.008*"questions" + '<br>'0.008*"experience" + 0.008*"program" + 0.008*"advice" + 0.007*"email" + ' '0.007*"hi" + 0.007*"first"' | Guidance |
| Topic 5 | '0.045*"health" + 0.043*"mental" + 0.023*"students" + 0.012*"online" + '<br>'0.009*"may" + 0.008*"please" + 0.007*"free" + 0.007*"us" + 0.007*"issues" + '<br>'0.007*"many" | Students |
| Topic 6 | 0.021*"f*****" + 0.015*"s***" + 0.013*"covid" + 0.011*"life" + '<br>'0.010*"group" + 0.009*"people" + 0.008*"bed" + 0.008*"found" + 0.007*"fact" '<br>'+ 0.007*"students"' | Covid |

*Table 5: the 2021data set common bi- and tri-grams*

| Common Bi-grams | ('mental', 'health'), 227), (('first', 'year'), 118), (('anyone', 'else'), 73), (('second', 'year'), 51), (('high', 'school'), 51), (('last', 'year'), 42), (('first', 'time'), 40), (('even', 'though'), 35), (('feel', 'free'), 32), (('social', 'anxiety'), 32) |
|---|---|
| Common-Trigrams | ('mental', 'health', 'issues'), 26), (('anyone', 'else', 'feel'), 11), (('serious', 'case', 'covid'), 11), (('anyone', 'else', 'feeling'), 10), (('would', 'greatly', 'appreciated'), 10), (('worried', 'catching', 'covid'), 10), (('catching', 'covid', 'feel'), 10), (('keep', 'risks', 'perspective'), 10), (('lose', 'sleep', 'prospect'), 10), (('sleep', 'prospect', 'catching'), 10) |

In the years 2022 and 2023, a sense of normalcy began to return to society at large and to the universities. With improved strategies for handling COVID-19, in-person lectures resumed, and campus life regained its usual rhythm. COVID-19 gradually became a less prominent concern. During this period, our expectation was that discussions about COVID-19 and mental health issues related to the pandemic would decrease. Examining the topics in 2022 reveals a notable reduction in references to anxiety among the keywords (Table 6). Many of these topics primarily center around university-related matters, indicating the return to regular university activities. The exception is Topic 4, which addresses mental health issues. Keywords such as depression, mental health, campus, and in-person may suggest concerns related to students' mental well-being as they transition back to in-person schooling. Given that students had been away for two years, it's reasonable to anticipate that some may encounter mental challenges upon their return.

Furthermore, while examining the frequently occurring bigrams and trigrams in 2022 (Table 7), we observed a further decrease in mentions of mental health compared to previous years, although it remains the top bigram. While mental health continues to be a prevalent topic of discussion, its association with COVID-19 and distance learning has notably diminished this year.

In 2023, there is a noticeable decline in discussions about anxiety and mental health (Table 8), similar to the trend observed in 2022. Interestingly, in Topic 0, the central theme revolves around sexual abuse, marking the first instance of such a specific focus. Additionally, in Topic 4, the presence of keywords like "mental," "health," "back," and "anxiety" suggests a connection to the apprehensions linked with returning to school. This may encompass concerns about academic performance, social interactions, or the general anxiety associated with the school environment. Moreover, upon analyzing the prevalent bigrams and trigrams for this year (Table 9), it's noteworthy that the term 'mental health' has the lowest occurrence, totaling 187, across the four years of our study. Interestingly, there are no other significant bigrams related to mental health or anxiety. The sole trigram suggesting a connection to mental health is 'can't help feel,' which appears 10 times in the text, implying a connection with individuals' emotions. Apart from that, the textual analysis indicates that discussions this year were more optimistic compared to those of the previous year.

*Table 6: the 2022 data topics*

| Topic 0 | '0.009*"questions" + 0.009*"email" + 0.009*"offer" + 0.009*"id" + '<br>'0.009*"resources" + 0.009*"ams" + 0.007*"exam" + 0.007*"check" + ' 0.007*"receive" +<br>0.007*"reach"' | University |
|---|---|---|
| Topic 1 | 0.020*"diagnosis" + 0.018*"club" + 0.015*"please" + 0.014*"potential" + "0.014*"selvesubc" + 0.012*"doctors" + 0.012*"us" + 0.011*"image" + ' 0.011*"service" + 0.010*"lately"') | Health care |
| Topic 2 | 0.013*"feel" + 0.013*"year" + 0.010*"would" + 0.010*"time" + 0.009*"people" '<br>'+ 0.008*"also" + 0.008*"first" + 0.008*"even" + 0.007*"want" + 0.007*"one"') | First year |
| Topic 3 | '0.026*"class" + 0.026*"course" + 0.022*"final" + 0.022*"exam" + '<br>'0.015*"grade" + 0.012*"prof" + 0.008*"posts" + 0.008*"transcript" + '<br>'0.007*"pass" + 0.007*"drop"' | University course |
| Topic 4 | '0.060*"health" + 0.059*"mental" + 0.019*"students" + 0.010*"services" + ' 0.008*"issues" +<br>0.007*"would" +<br>0.007*"depression" + 0.007*"campus" + ' 0.006*"classes" + 0.006*"inperson"') | Mental health |
| Topic 5 | '0.021*"courses" + 0.019*"course" + 0.011*"would" + 0.010*"science" + '<br>'0.008*"take" + 0.008*"major" + 0.007*"cs" + 0.007*"program" + 0.006*"weeks" '<br>'+ 0.006*"body"') | University courses |
| Topic 6 | '0.013*"ubyssey" + 0.012*"httpspreview" + 0.010*"free" + 0.008*"qualtrics" + '<br>'0.008*"nlm" + 0.008*"daniel" + 0.008*"unofficial" + 0.008*"research" + ' 0.007*"university" +<br>0.006*"survey"') | Undetermined |

*Table 7: the 2022 data set common bi- and tri-grams*

| Common Bi-grams | ('mental', 'health'), 229), (('first', 'year'), 106), (('anyone', 'else'), 62), (('last', 'year'), 49), (('high', 'school'), 44), (('first', 'time'), 38), (('final', 'exam'), 35), (('hi', 'everyone'), 34), (('even', 'though'), 32), (('health', 'issues'), 32) |
|---|---|
| Common-Trigrams | [(('mental', 'health', 'issues'), 29), (('please', 'use', 'thread'), 20), (('use', 'thread', 'discuss'), 19), (('thread', 'discuss', 'mentalhealth'), 19), (('discuss', 'mentalhealth', 'related'), 19), (('mentalhealth', 'related', 'topics'), 19), (('related', 'topics', 'includes'), 19), (('topics', 'includes', 'personal'), 19), (('includes', 'personal', 'venting'), 19), (('personal', 'venting', 'strategies'), 19)] |

*Table 8: The 2023 data topics*

| Topic 0 | '0.020*"students" + 0.013*"ams" + 0.009*"health" + 0.009*"sexual" + ' '0.008*"community" + 0.007*"social" + 0.007*"clause" + 0.007*"people" + ' '0.006*"seeing" + 0.006*"assault"') | Abuse |
|---|---|---|
| Topic 1 | '0.022*"yet" + 0.008*"unit" + 0.008*"st" + 0.008*"hasnt" + 0.007*"fresh" + ' '0.007*"helping" + 0.006*"friday" + 0.005*"book" + 0.005*"httpspreview" + ' '0.005*"submit"' | Undetermined |
| Topic 2 | '0.017*"study" + 0.015*"policy" + 0.013*"please" + 0.012*"research" + ' '0.011*"students" + 0.010*"ca" + 0.008*"completely" + 0.008*"us" + ' '0.007*"team" + 0.007*"receive"') | Academia |
| Topic 3 | '0.015*"applying" + 0.012*"international" + 0.012*"application" + ' '0.011*"psych" + 0.010*"tuition" + 0.010*"related" + 0.010*"credits" + ' '0.009*"gone" + 0.008*"side" + 0.008*"approved"') | University program |
| Topic 4 | '0.014*"health" + 0.013*"time" + 0.011*"mental" + 0.010*"back" + ' '0.010*"even" + 0.009*"anxiety" + 0.009*"feel" + 0.008*"would" + 0.008*"one" ' '+ 0.007*"last"') | Mental health |
| Topic 5 | '0.014*"offer" + 0.013*"academic" + 0.012*"food" + 0.012*"email" + ' '0.010*"prof" + 0.009*"supposed" + 0.009*"complete" + 0.009*"got" + "0.009*"form" + 0.008*"entire"') | Academia |
| Topic 6 | '0.019*"year" + 0.011*"also" + 0.010*"school" + 0.009*"would" + ' '0.009*"people" + 0.009*"want" + 0.009*"course" + 0.008*"feel" + ' '0.008*"anyone" + 0.007*"first"') | First year |

*Table 9: the 2023 data set common bi- and tri-grams*

| Common Bi-grams | [(('mental', 'health'), 187), (('first', 'year'), 166), (('anyone', 'else'), 71), (('even', 'though'), 65), (('grad', 'school'), 53), (('last', 'year'), 45), (('high', 'school'), 41), (('social', 'justice'), 40), (('second', 'year'), 39), (('wondering', 'anyone'), 37)] |
|---|---|
| Common-Trigrams | [(('would', 'greatly', 'appreciated'), 13), (('view', 'pollhttpswww', 'compoll'), 12), (('certain', 'progressive', 'ideas'), 12), (('free', 'speech', 'campuses'), 12), (('ymca', 'greater', 'toronto'), 12), (('substance', 'use', 'amnesty'), 11), (('use', 'amnesty', 'clause'), 11), (('please', 'feel', 'free'), 10), (('cant', 'help', 'feel'), 10), (('overnight', 'mobile', 'team'), 10)] |

Upon examining the generated topics for each university, numerous similarities become apparent. Firstly, all universities exhibit at least three topics that pertain to mental health and anxiety. This observation holds significant weight, aligning with findings from a study conducted by the Canadian Alliance of Student Associations[10], where three-quarters of students reported experiencing adverse mental health effects during the 2021-2022 academic years. The Alliance is composed from student associations across Canada. It consists of 24 member associations that collectively represent more than 275,000 post-secondary students. Given the prevalence of negative sentiments within two years of our study, we anticipated their reflection in our topic representation. Furthermore, additional topics emerge that center around university grades and associated struggles. This connection is noteworthy as poor academic performance is directly correlated with low mental health and anxiety (Alghoul, 2022).

The following Tables 10 – 13 report results of topic modelling per university.

---

[10] Canadian Alliance of Student Associations (casa-acae.com)

*Table 10: Topics of the Waterloo data.*

| Topic 0 | '0.065*"please" + 0.061*"thread" + 0.052*"subreddit" + 0.051*"welcome" + '<br>'0.051*"constructive" + 0.048*"use" + 0.041*"note" + 0.019*"nothing" + '<br>'0.019*"rest" + 0.012*"discuss"') | Reddit |
|---|---|---|
| Topic 1 | '0.011*"wednesday" + 0.009*"ece" + 0.007*"work" + 0.006*"due" + '<br>'0.006*"water" + 0.006*"applying" + 0.006*"stat" + 0.006*"deal" + '<br>'0.005*"behind" + 0.005*"profs"' | Undetermined. |
| Topic 2 | '0.019*"pop" + 0.014*"wusa" + 0.008*"students" + 0.005*"geese" + '<br>'0.005*"university" + 0.005*"governance" + 0.004*"support" + '<br>'0.004*"administration" + 0.004*"fellow" + 0.004*"threat"' | University program |
| Topic 3 | '0.023*"mental" + 0.021*"health" + 0.018*"university" + 0.018*"might" + '<br>'0.018*"anything" + 0.017*"life" + 0.017*"advice" + 0.017*"please" + '<br>'0.016*"keep" + 0.016*"similar"' | Mental health |
| Topic 4 | '0.014*"term" + 0.013*"coop" + 0.012*"anyone" + 0.012*"would" + '<br>'0.010*"first" + 0.010*"feel" + 0.010*"year" + 0.009*"since" + '<br>'0.009*"anxious" + 0.008*"time"' | School anxiety |
| Topic 5 | '0.020*"struggling" + 0.016*"counselling" + 0.012*"set" + 0.010*"health" + '<br>'0.007*"adhd" + 0.007*"services" + 0.006*"sugar" + 0.006*"options" + '<br>'0.005*"deadline" + 0.004*"psychologist"' | Mental health |
| Topic 6 | '0.032*"others" + 0.031*"comments" + 0.017*"feel" + 0.017*"people" + '<br>'0.015*"anxiety" + 0.009*"friends" + 0.008*"even" + 0.008*"always" + '<br>'0.007*"students" + 0.007*"make"') | Anxiety |

*Table 11: Topics of the McGill data.*

| | | |
|---|---|---|
| Topic 0 | '0.017*"alone" + 0.013*"anxiety" + 0.010*"eating" + 0.009*"disorder" + "0.009*"thoughts" + 0.009*"worse" + 0.008*"usually + 0.008*"personal" + ' 0.007*"lost" + 0.006*"every"' | Mental health |
| Topic 1 | '0.013*"ideas" + 0.009*"students" + 0.007*"justice" + 0.006*"progressive" + ' '0.006*"french" + 0.005*"many" + 0.005*"instagram" + 0.005*"google" + ' '0.005*"mcgillians" + 0.004*"post"' | Social justice |
| Topic 2 | "0.014*"please" + 0.013*"students" + 0.012*"university" + 0.012*"issues" + ' '0.012*"us" + 0.009*"new" + 0.007*"questions" + 0.007*"support" + ' '0.007*"back" + 0.006*"international"' | University reform |
| Topic 3 | '0.021*"comp" + 0.018*"courses" + 0.017*"research" + 0.013*"math" + ' '0.013*"ssmu" + 0.011*"taken" + 0.011*"program" + 0.008*"elective" + ' '0.008*"course" + 0.008*"project"' | Academia |
| Topic 4 | "0.027*"covid" + 0.023*"students" + 0.018*"health" + 0.015*"vaccinated" + ' '0.009*"mental" + 0.009*"learning" + 0.008*"cases" + 0.006*"serious" + ' '0.006*"com" + 0.005*"risk"' | Covid |
| Topic 5 | "0.013*"semester" + 0.012*"class" + 0.012*"anyone" + 0.010*"exam" + ' '0.010*"course" + 0.010*"time" + 0.009*"mental" + 0.008*"first" + ' '0.008*"would" + 0.008*"year"' | School struggles |
| Topic 6 | '0.016*"feel" + 0.013*"people" + 0.012*"even" + 0.011*"would" + ' '0.011*"health" + 0.011*"want" + 0.009*"go" + 0.008*"also" + 0.008*"mental" ' '+ 0.008*"going"') | Mental health |

*Table 12: The UBC data topics*

| Topic 0 | '0.013*"vancouver" + 0.008*"heard" + 0.007*"faculty" + 0.007*"posts" + ' '0.007*"international" + 0.006*"would" + 0.006*"groups" + 0.006*"canada" + ' '0.005*"support" + 0.005*"us"' | Undetermined |
|---|---|---|
| Topic 1 | 0.015*"students" + 0.012*"questions" + 0.011*"health" + 0.011*"issues" + ' '0.010*"therapy" + 0.010*"counselling" + 0.010*"support" + 0.009*"also" + ' '0.009*"services" + 0.007*"resources"' | Mental health |
| Topic 2 | '0.020*"please" + 0.017*"study" + 0.015*"ca" + 0.010*"said" + 0.009*"may" + ' '0.008*"email" + 0.007*"sessions" + 0.007*"free" + 0.006*"note" + ' '0.006*"club"' | ? |
| Topic 3 | '0.027*"ams" + 0.020*"coverage" + 0.013*"reading" + 0.009*"doctor" + ' '0.008*"nd" + 0.007*"pngwidth" + 0.006*"ubyssey" + 0.006*"asians" + ' '0.005*"grateful" + 0.005*"afford"' | Health |
| Topic 4 | '0.013*"course" + 0.011*"time" + 0.011*"people" + 0.010*"exam" + ' '0.010*"feel" + 0.007*"final" + 0.007*"going" + 0.007*"well" + 0.007*"see" + ' '0.007*"exams"' | Academia |
| Topic 5 | '0.016*"mental" + 0.015*"health" + 0.015*"year" + 0.011*"feel" + ' '0.009*"even" + 0.008*"first" + 0.008*"would" + 0.008*"help" + 0.008*"time" ' '+ 0.007*"go"' | Mental health |
| Topic 6 | '0.025*"students" + 0.013*"mar" + 0.010*"diagnosis" + 0.009*"concerns" + ' '0.009*"community" + 0.007*"lack" + 0.007*"inperson" + 0.005*"chat" + ' '0.005*"sick" + 0.005*"people"') | Covid |

*Table 13: The UofT data topics.*

| Topic 0 | '0.014*"go" + 0.014*"would" + 0.013*"school" + 0.010*"also" + 0.009*"want" + ' '0.009*"well" + 0.009*"good" + 0.008*"university" + 0.008*"experience" + ' '0.007*"better"' | University career |
|---|---|---|
| Topic 1 | '0.010*"university" + 0.010*"interview" + 0.009*"toronto" + 0.009*"rn" + ' '0.007*"test" + 0.006*"service" + 0.006*"offers" + 0.006*"concepts" + ' '0.006*"certificate" + 0.006*"police"' | ? |
| Topic 2 | '0.016*"grade" + 0.016*"year" + 0.016*"gpa" + 0.014*"without" + ' '0.014*"science" + 0.013*"major" + 0.012*"program" + 0.012*"school" + ' '0.011*"average" + 0.010*"grades"' | University grades |
| Topic 3 | '0.021*"feel" + 0.016*"even" + 0.016*"people" + 0.014*"anxiety" + ' '0.010*"time" + 0.009*"cant" + 0.009*"help" + 0.009*"work" + 0.008*"want" + ' '0.007*"many"' | Anxiety |
| Topic 4 | '0.016*"mat" + 0.011*"needed" + 0.009*"pm" + 0.008*"effort" + 0.007*"note" + ' '0.007*"eco" + 0.007*"nervous" + 0.007*"midterm" + 0.007*"continue" + ' '0.007*"today"' | Anxiety |
| Topic 5 | '0.038*"health" + 0.035*"mental" + 0.013*"help" + 0.012*"exam" + ' '0.012*"anxiety" + 0.011*"need" + 0.009*"students" + 0.008*"services" + ' '0.007*"prof" + 0.007*"program"' | Mental health |
| Topic 6 | '0.031*"year" + 0.022*"anyone" + 0.019*"course" + 0.016*"first" + ' '0.015*"semester" + 0.015*"would" + 0.014*"got" + 0.014*"courses" + ' '0.013*"still" + 0.012*"back"' | First year |

## 5    Discussion

As we said early on, we worked with the data pulled from subreddits of the top universities in Canada. Being enrolled in the top universities, students may have better access to scholarships and bursaries if compared with mid- and low ranked universities; this observation is especially true for Waterloo that holds the highest rank in scholarships and bursaries. Graduates of top universities may have better job prospects too. Our selection of the four universities may influence the obtained results. It could contribute to a significant presence of positive posts in our data. Moreover, the dataset itself is predominantly composed of positive topics. This prevalence of positivity within the dataset significantly shapes the topic modeling analysis.

Our investigation affirms that during the 2020-2021 timeframe, a heightened occurrence of mental health issues and discussions on anxiety emerged, largely attributed to the impact of the COVID-19 pandemic. Secondly, as we anticipated, the years 2022-2023 demonstrated a decline in conversations about mental health, aligning with our expectation of a shift toward a more normalized state of affairs.

## 6    Conclusions and Future Work

We analyzed textual data from subreddits of four prominent Canadian universities (Waterloo, McGill, UofT, and UBC). The data spanned the years 2020-2023. We used year-long intervals for the longitudinal aspect of the study. Our empirical investigation involved a year-by-year analysis, focusing on calculating sentiments for each post and constructing machine learning models to effectively classify sentiments as positive, negative, or neutral. Additionally, we implemented an LDA topic model to extract deeper insights from the text and understand its interrelations. Throughout these processes, we formulated hypotheses, some of which were confirmed while others were disproven.

One hypothesis that held true was the surge in COVID-related mental health and anxiety between 2020-2021, followed by a decline in 2022-2023. Our textual analysis, including the construction of topics and examination of frequent bigrams and trigrams, aligned with this hypothesis, possibly reflecting the increased occurrence of positive terms during those years.

Our assumption was that the dataset, derived from generally negative keywords, would predominantly consist of negative sentiment posts. However, positive sentiments held a substantial presence in the data. This can be attributed to the significant echo-chamber effect of social networks (Terren, Borge-Bravo, 2021). The effect stipulates that online community members usually exhibit similar sentiments when posting online. At the same time, they could express different sentiments in private communications, thus leading to divergence of our results to the survey results reported by Abdi (2023) and Panchal et al (2021). We envision that future work can focus on subreddits of mid-rank and low rank universities.